\documentclass[a4paper]{spie}  

\usepackage{amsmath,amsfonts,amssymb}
\usepackage{graphicx}
\usepackage[colorlinks=true, allcolors=blue]{hyperref}

\title{Kyoto's Event-Driven X-ray Astronomy SOI pixel sensor for the FORCE mission}

\author[a]{Takeshi G. Tsuru}
\author[a]{Hideki Hayashi}
\author[a]{Katsuhiro Tachibana}
\author[a]{Sodai Harada}
\author[a]{Hiroyuki Uchida}
\author[a]{Takaaki Tanaka}
\author[b]{Yasuo Arai}
\author[b]{Ikuo Kurachi}
\author[c]{Koji Mori}
\author[c]{Ayaki Takeda}
\author[c]{Yusuke Nishioka}
\author[c]{Nobuaki Takebayashi}
\author[c]{Shoma Yokoyama}
\author[c]{Kohei Fukuda}
\author[d]{Takayoshi Kohmura}
\author[d]{Kouichi Hagino}
\author[d]{Kenji Ohno}
\author[d]{Kohsuke Negishi}
\author[d]{Keigo Yarita}
\author[e]{Shoji Kawahito}
\author[e]{Keiichiro Kagawa}
\author[e]{Keita Yasutomi}
\author[e]{Sumeet Shrestha}
\author[e]{Shunta Nakanishi}
\author[f]{Hiroki Kamehama}
\author[g]{Hideaki Matsumura}

\affil[a]{Department of Physics, Graduate School of Science, Kyoto University, Kitashirakawa, Sakyo-ku, Kyoto, 606-8502, Japan}
\affil[b]{Institute of Particle and Nuclear Studies, High Energy Accelerator Research Org., KEK, 1-1 Oho, Tsukuba 305-0801, Japan}
\affil[c]{Department of Applied Physics and Electronic Engineering, Faculty of Engineering, University of Miyazaki, 
1-1 Gakuen Kibanadai-Nishi, Miyazaki, 889-2192, Japan}
\affil[d]{Department of Physics, Faculty of Science and Technology, Tokyo University of Science, 2641 Yamazaki, Noda, Chiba 278-8510, Japan}
\affil[e]{Research Institute  of Electronics, Shizuoka University, Johoku 3-5-1, Naka-ku, Hamamatsu 432-8011, Japan}
\affil[f]{Information and Communication Systems Engineering, Okinawa National College of Technology, Henoko 905, Nago, Okinawa, Japan}
\affil[g]{Kavli IPMU, The University of Tokyo, 5-1-5 Kashiwanoha, Kashiwa, Chiba, 277-8583, Japan}

\authorinfo{Further author information: (Send correspondence to T.G.T.)\\T.G.T.: E-mail: tsuru@cr.scphys.kyoto-u.ac.jp, Telephone: 81 75 753 3868}

\begin{document} 
\maketitle


\begin{abstract} 
We have been developing monolithic active pixel sensors, X-ray Astronomy SOI pixel sensors, XRPIXs, 
based on a Silicon-On-Insulator (SOI) CMOS technology as soft X-ray sensors 
for a future Japanese mission, FORCE (Focusing  On Relativistic universe and Cosmic Evolution). 
The mission is characterized by broadband (1-80 keV) X-ray imaging spectroscopy with high angular resolution ($<15$~arcsec), 
with which we can achieve about ten times higher sensitivity in comparison to the previous missions above 10~keV. 
Immediate readout of only those pixels hit by an X-ray is available by an event trigger output function implemented in each pixel with the time resolution higher than $10~{\rm \mu sec}$ (Event-Driven readout mode). 
It allows us to do fast timing observation and also reduces non-X-ray background dominating at a high X-ray energy band above 5--10~keV 
by adopting an anti-coincidence technique. 
In this paper, we introduce our latest results from the developments of the XRPIXs. 
(1) We successfully developed a 3-side buttable back-side illumination device 
with an imaging area size of 21.9~mm$\times$13.8~mm and an pixel size of $36~{\rm \mu m} \times 36~{\rm \mu m}$. 
The X-ray throughput with the device reaches higher than 0.57~kHz in the Event-Driven readout mode. 
(2) We developed a device using the double SOI structure and found that the structure improves the spectral performance in the Event-Driven readout mode by suppressing the capacitive coupling interference between the sensor and circuit layers.
(3) We also developed a new device equipped with the Pinned Depleted Diode structure and confirmed that the structure reduces the dark current generated at the interface region between the sensor and the SiO$_2$ insulator layers. 
The device shows an energy resolution of 216~eV in FWHM at 6.4~keV in the Event-Driven readout mode. 

%
%
\end{abstract}


\section{INTRODUCTION: the FORCE mission}
\label{sec:intro}  


We are now proposing a future Japan-lead X-ray mission, FORCE, to be realized in 2026\cite{Mori:2016ju, Nakazawa:2018iy}. 
The primary scientific objective of the FORCE mission is to hunt for ``missing black holes'' in various mass-scales and to trace their cosmic evolution. 
The black holes include super-massive black holes in highly-obscured AGNs, 
intermediate massive black holes which might be the seeds for super-massive black holes, and isolated stellar mass black holes. 
We identify and quantify families of black holes whose populations and evolutions are unknown.
Through this, we explore the evolution paths of the Universe. 

Broadband observation is essential to uncover buried AGNs, to measure the masses of the black holes and to distinguish isolated stellar mass black holes from neutron stars and white dwarfs. 
The FORCE mission offers broadband (1-80~keV) X-ray imaging spectroscopy with high angular resolution $<15$~arcsec in HPD, 
which offers about ten times higher sensitivity than NuSTAR at 1-80~keV\cite{Harrison:2013iq}. 
For this purpose, FORCE has 3 identical pairs of super-mirrors and detectors. 
X-ray mirror is Light-weight Si mirror provided by NASA/GSFC\cite{Zhang:2018cc}. 
We adopt a hybrid type of detector to cover the broadband, whose overall design is based on the HXI onboard the Hitomi satellite \cite{Nakazawa:2018gc}.  
The detector imager consists of Si and CdTe layers. 
In this report, we introduce the latest results from the development of X-ray SOI pixel sensors as the Si layer. 

%
%

\section{XRPIX: Event Driven X-ray SOI Pixel Sensor}
\label{sec:SOIPIX}  
Reduction of non-X-ray background (NXB) due to interactions of cosmic-rays is essential to detect faint X-ray objects, 
especially in the energy band above 10~keV. 
We adopt the anti-coincidence technique with scintillators surrounding the X-ray imager to reduce NXB. 
Since the counting rate of the scintillators is estimated to be around $\sim 10$~kHz in a low earth orbit\cite{Takahashi:2007ev, Kokubun:2007is}, the time resolution of X-ray CCD is too poor to use this technique. 
On the other hand, the double-sided Si strip detector (DSSD) can not detect X-rays below $\sim 2$~keV 
due to its relatively high readout noise, 
although the time resolution is high enough to adopt the anti-coincidence technique. 
This is the motivation for developing the X-ray astronomy SOI pixel sensors. 

The SOI pixel sensor is monolithic, using a bonded wafer of a high resistivity depleted Si layer for X-ray detection, 
a SiO$_2$ insulator layer (buried oxide: BOX) and a low resistivity Si layer for CMOS circuits. 
``XRPIXs'' refer to the X-ray astronomy SOI pixel sensors. 
In order to realize the time resolution significantly higher than $\sim 100~{\rm \mu sec}$ ($1/10$~kHz), 
each pixel of an XRPIX is equipped with its own trigger logic circuit and analogue readout CMOS circuit, 
which is the most important feature of the XRPIX. 
Table~\ref{tab:fonts} shows the target specification of the XRPIX for the FORCE mission. 

\begin{table}[b]
\caption{\label{tab:fonts}Target Specification of the XRPIX}
\vspace{-0.25cm}
\begin{center}
\begin{tabular}{ll}
\hline
\rule[-1ex]{0pt}{3.5ex}  Item & Specification  \\
\hline
\rule[-1ex]{0pt}{3.5ex}	Imaging    & Imaging Area: $15\times 45~{\rm mm^2}$, Pixel Size: 36~${\rm \mu m}$ ($0.74''$ at F$=10$~m)\\
\rule[-1ex]{0pt}{3.5ex}	Bandpass   & 1--40~keV (Requirement), 0.3--40~keV (Goal) \\
\rule[-1ex]{0pt}{3.5ex}	\ \ Back side dead layer thickness & $1~{\rm \mu m}$ (Requirement), $0.1~{\rm \mu m}$ (Goal) \\
\rule[-1ex]{0pt}{3.5ex}	\ \ Depletion layer thickness & $200~{\rm \mu m}$ (Requirement), $500~{\rm \mu m}$ (Goal) \\
\rule[-1ex]{0pt}{3.5ex}	Energy resolution at 6~keV (FWHM) & $300~{\rm eV}$ (Requirement),  $140~{\rm eV}$ (Goal) \\
\rule[-1ex]{0pt}{3.5ex}	Equivalent noise charge (rms) & $10~{\rm e}$ (Requirement), $3~{\rm e}$ (Goal) \\
\rule[-1ex]{0pt}{3.5ex}	Time resolution & $10~{\rm \mu sec}$ \\
\rule[-1ex]{0pt}{3.5ex}	Throughput & $2~{\rm kHz}$ (counting rate of the Crab nebula) \\
\rule[-1ex]{0pt}{3.5ex} Non-X-ray background at 20~keV & $1/100$ of X-ray CCD
						(${\rm 5\times 10^{-5}~cps/keV/10\times 10mm^{2}}$) \\
\hline
\end{tabular}
\end{center}
\end{table}

\begin{figure} [ht]
	\begin{center}
	\begin{tabular}{c} 
	\includegraphics[width=12cm]{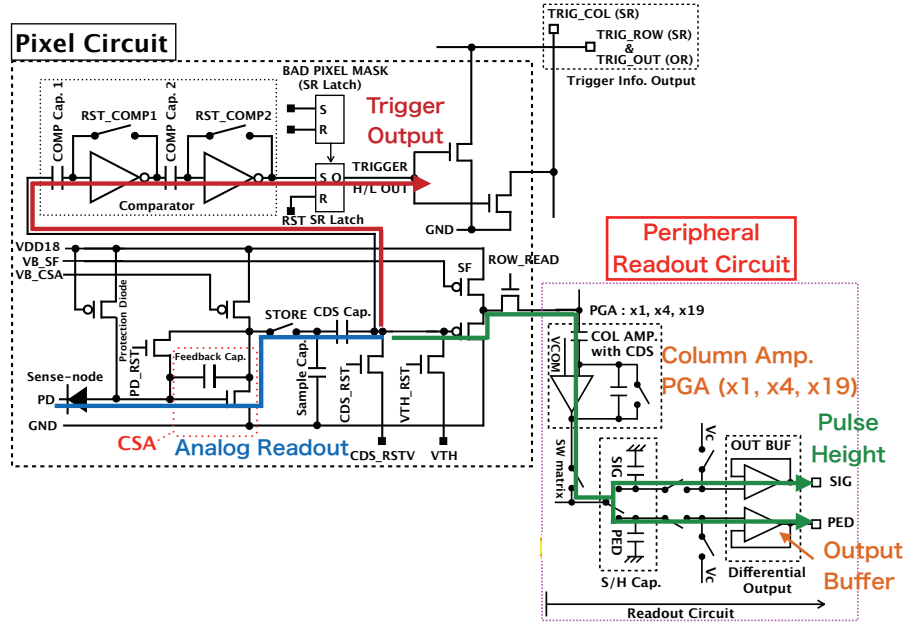}
	\end{tabular}
	\end{center}
	\caption[example] 
	{ \label{fig:PixelPeropheralCirc} 
	In-pixel CMOS circuit and peripheral readout circuit of XRPIX5b having an n-type sensor layer. 
	}
\end{figure}

Figure~\ref{fig:PixelPeropheralCirc} shows the in-pixel CMOS circuit implemented in each pixel and the peripheral readout circuit. 
The in-pixel CMOS circuitry consists of an analog readout circuit and an trigger output circuit. 
The signal charge from the PN diode is converted to signal voltage at the charge sensitive amplifier (CSA). 
The correlated double sampling (CDS) circuit, consisting of the CDA capacitance and CDS reset transistor, 
cancels the reset noise generated at the CSA circuit. 
The signal voltage is read out through the source follower (SF) implemented in the pixel, 
and the peripheral readout circuit consisting of the column programable gain amplifier and output buffer. 
We use an inverter-chopper type of comparator in the trigger output circuit. 
The threshold level is set at VTH through the VTH reset transistor. 
The trigger output signal can be masked at pixel level. 
We read out $8\times 8$ pixels for each event to see the charge sharing 
and to collect all the signal charge produced by an X-ray event. 

\begin{figure} [ht]
	\begin{center}
	\begin{tabular}{c} 
	\includegraphics[height=5cm]{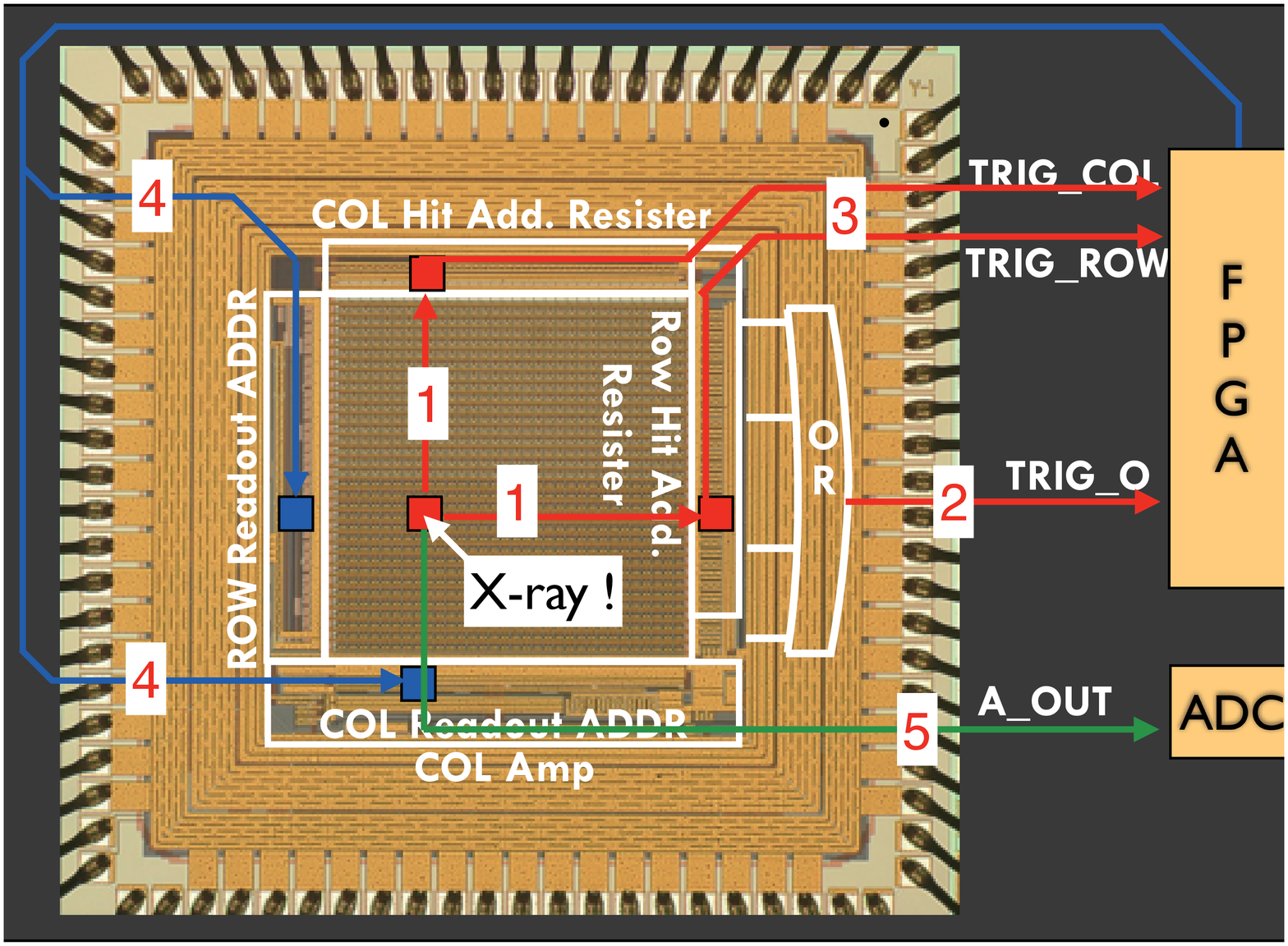}
	\end{tabular}
	\end{center}
	\caption[example] 
	{ \label{fig:EventDrivenSequence} 
	The readout sequence in the Event-Driven readout mode. 
	}
\end{figure}

The XRPIX has two readout modes. 
One is ``Frame readout mode'', in which all the pixels are read out serially like a CCD without using the trigger function. 
The other is ``Event-Driven readout mode'', in which the hit pixel and its surrounding $8\times 8$ pixels, 
with the hit pixel in the center, are read out using the trigger function. 
Figure~\ref{fig:EventDrivenSequence} shows how to read out an X-ray event in the Event-Driven readout mode. 
(1) Assuming an X-ray is detected at the pixel labeled with ``X-ray !'' in Figure~\ref{fig:EventDrivenSequence}, 
the pixel trigger circuit writes its address into the row and column hit resisters. 
(2) A trigger flag (TRIG\_O) is immediately sent to the FPGA.
(3) The FPGA reads the address of the hit pixel. 
(4) The FPGA writes the address of the pixel to be read into the row and column readout registers, 
(5) then reads out the analog signal of the pixel with the ADC. 

\section{Results from recent developments}
\label{sec:Results}  

\subsection{Imaging in the Event Driven Readout Mode}
XRPIX5b is a 3-side buttable back-illumination device and the largest XRPIX sensor we have processed. 
The device has an imaging area size of 21.9~mm$\times$13.8~mm, an pixel size of $36~{\rm \mu m} \times 36~{\rm \mu m}$ and a format of $608\times 384$. 
The sensor layer is n-type Si with a resistivity of $\sim 5~{\rm \Omega cm}$. 

\begin{figure} [ht]
	\begin{center}
	\begin{tabular}{c} 
	\includegraphics[width=8cm]{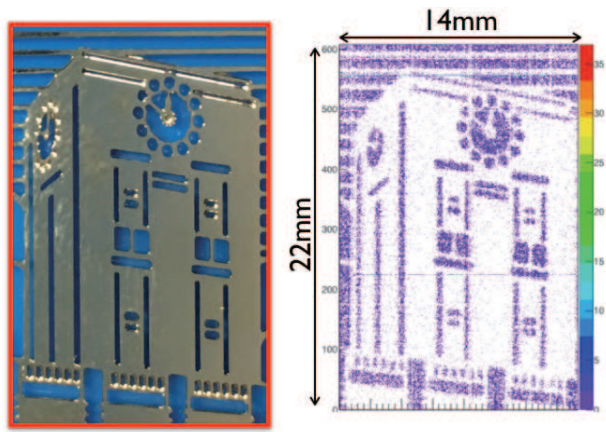}
	\end{tabular}
	\end{center}
	\caption[example] 
	{ \label{fig:EventDrivenImaging} 
	Demonstration of X-ray imaging performance of XRPIX5b in the Event-Driven readout mode. 
	The right panel shows a Cd-109 X-ray image obtained with XRPIX5b in the Event-Driven readout at room temperature.
	The left panel is a photo of the openwork used in the demonstration as the mask (the clock tower of Kyoto university). 
	}
\end{figure}

Figure~\ref{fig:EventDrivenImaging} shows a demonstration of a Cd-109 X-ray image obtained with XRPIX5b operated in the Event-Driven readout mode\cite{Hayashi:2018xx}. 
We placed the openwork of the clock tower of Kyoto University as the mask over the XRPIX5b, 
and applied a back bias voltage of 10~V at room temperature. 
The thickness of the depletion layer is estimated to be $\sim 130~{\rm \mu m}$. 
Also, the device successfully detected Am-241 X-rays at a count rate of $\sim 570~{\rm Hz}$ without using the mask. 
The back bias voltage was 10~V and the operating temperature was $-60{\rm ^\circ C}$.
Since the count rate was limited by the RI source used in the experiment, the maximum throughput would be higher than $\sim 570~{\rm Hz}$. 


\subsection{Double-SOI structure}
\begin{figure} [ht]
	\begin{center}
	\begin{tabular}{c} 
	\includegraphics[width=14cm]{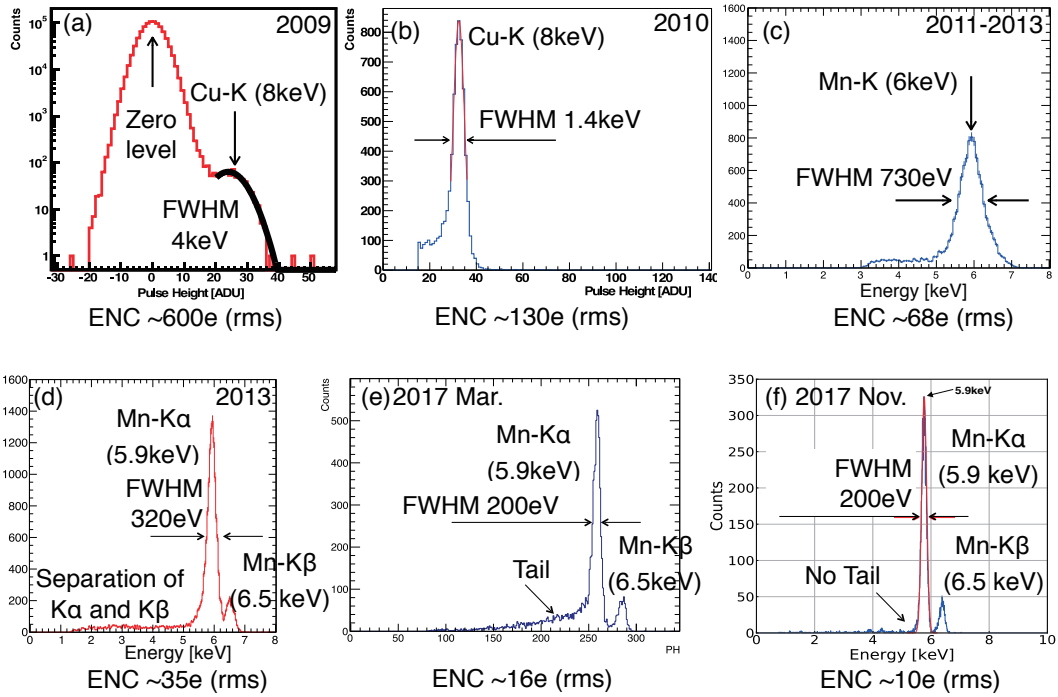}
	\end{tabular}
	\end{center}
	\caption[example] 
	{ \label{fig:HistSpectralPerformance} 
	Evolution of the spectral performance of X-ray astronomy SOIPIX 
	in the Frame readout\cite{Ryu:2011ey,Takeda:2015fk,Kamehama:2017kp,Takeda:2018xx}. 
	}
\end{figure}

Figure~\ref{fig:HistSpectralPerformance} shows the evolution of the spectral performance in the Frame readout mode from 2009 to 2017. 
The spectral performance in the Frame readout mode became significantly better by reducing the parasitic capacitance at the sense node, by introducing the charge sensitive amplifier instead of the source follower and by developing new device structures. 

\begin{figure} [ht]
	\begin{center}
	\begin{tabular}{c} 
	\includegraphics[width=14cm]{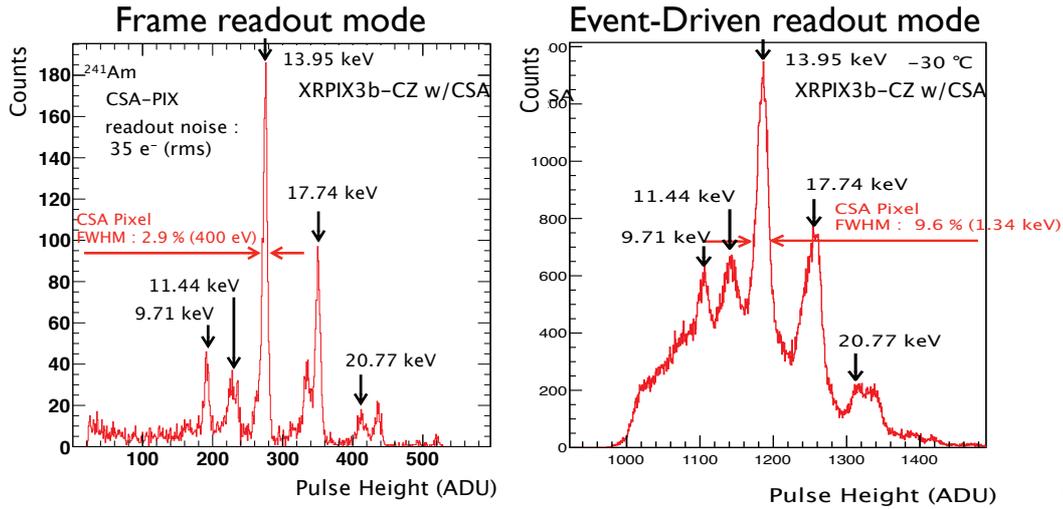}
	\end{tabular}
	\end{center}
	\caption[example] 
	{ \label{fig:XRPIX3b_Frame_EventDriven_Spec} 
	The spectra of the Am-241 X-rays obtained with an XRPIX3b (left) in the Frame readout mode and 
	(right) in the Event-Driven readout mode\cite{Takeda:2015fk}. 
	}
\end{figure}

Figure~\ref{fig:XRPIX3b_Frame_EventDriven_Spec} shows the spectra obtained with an identical device of XRPIX3b in the Frame and Event-Driven modes, respectively\cite{Takeda:2015fk}. 
The spectral performance in the Event-Driven readout mode was significantly worse than that in the Frame readout mode\cite{Takeda:2014uu}. 
In the Event-Driven readout mode, the energy resolution is poor and there is an offset in the output channel. 
The difference between the two modes is whether the in-pixel digital circuit is used or not. 
Therefore, it is conceivable that the interference with the analog signal due to the operation of the digital circuit degrades the spectral performance in the Event-Driven readout mode．

\begin{figure} [ht]
	\begin{center}
	\begin{tabular}{c} 
	\includegraphics[width=14cm]{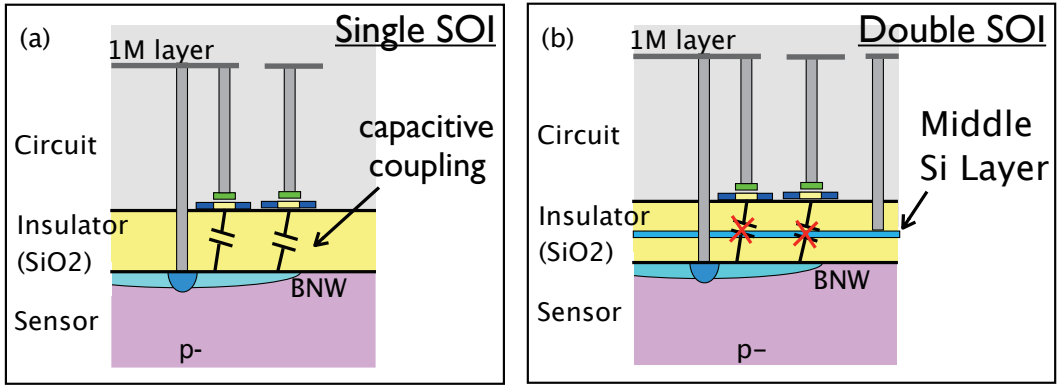}
	\end{tabular}
	\end{center}
	\caption[example] 
	{ \label{fig:DoubleSOI_Structure} 
	Cross-sectional views of (a) the single SOI and (b) Double-SOI structures\cite{Miyoshi:2013ir,Honda:2014te,Hara:2015ub}. 
	A p-type Si sensor layer is assumed in this figure. 
	}
\end{figure}

We found two causes degrading the spectral performance in the Event-Driven readout mode. 
First is that the analog and digital circuits have a common power supply line, resulting in the common impedance coupling. 
We already modified the power lines in the latest device so that each circuit has its own designated power line.  
The other is crosstalk between the digital circuit and the buried well (n-type buried well is applied for the p-type Si sensor layer), which is electrically connected to the sense-node. 
Figure~\ref{fig:DoubleSOI_Structure} (a) shows the cross-sectional views of the single SOI structure, which we have been using so far. 
There is the capacitive coupling between the BNW (buried n-well) and the trigger signal line in the circuit layer\cite{Takeda:2014uu}.

In order to solve the interference problem, we adopt a Double-SOI wafer (DSOI) in which we introduce an additional Si layer (middle Si layer) between the circuit and sensor layers\cite{Ohmura:2016ki}. 
The BOX layer is interleaved with the middle Si layer\cite{Miyoshi:2013ir,Honda:2014te,Hara:2015ub}, 
as shown in Figure~\ref{fig:DoubleSOI_Structure} (b). 
Note that the sensor layer of this Double-SOI device is p-type whereas that of the XRPIX5b is n-type. 
The middle Si layer is expected to act as an electrostatic shield and to reduce the capacitive coupling between the BNW and the digital circuit.


\begin{figure} [ht]
	\begin{center}
	\begin{tabular}{c} 
	\includegraphics[width=14cm]{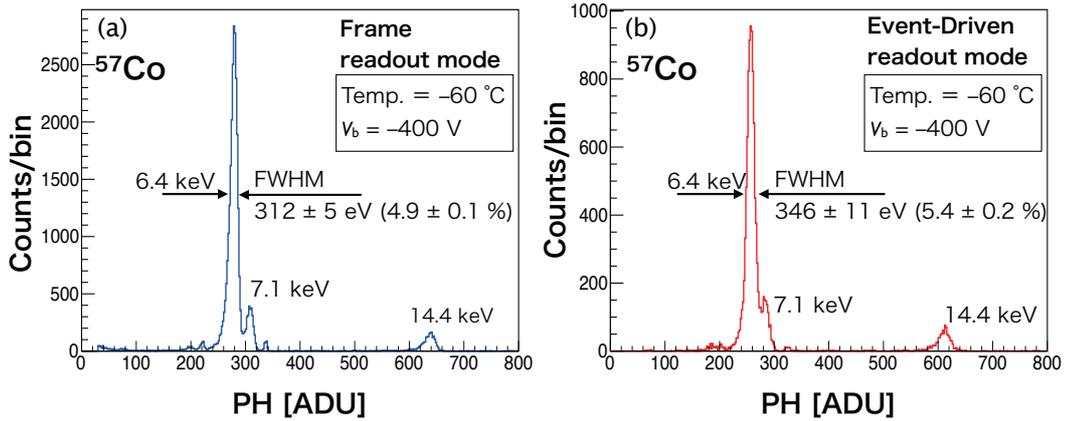}
	\end{tabular}
	\end{center}
	\caption[example] 
	{ \label{fig:XRPIX6D-PCZ-FI-300um_Frame_EventDriven} 
	Co-57 spectra obtained with the Double-SOI device (XRPIX6D) 
	(a) in the Frame readout mode and (b) in the Event-Driven readout mode\cite{Takeda:2018xx}. 
	}
\end{figure}

Figure~\ref{fig:XRPIX6D-PCZ-FI-300um_Frame_EventDriven} shows the spectra of Co-57 X-ray we obtained with an XRPIX6D having the Double-SOI structure in the Frame and Event-Driven readout modes. 
The performance in the Event-Driven mode is significantly improved by adopting the Double-SOI structure 
and is now close to that in the Frame readout mode. 
The energy resolutions are 312~eV and 346~eV at 6~keV in the Frame and Event-Driven readout modes, respectively. 
No significant offset in the output channel is observed in the Event-Driven readout mode. 
The results show that the crosstalk between the circuit and sensor layers is suppressed as expected. 
We found that the sense-node gain is increased by about a factor of two in comparison to the single SOI device having the same design of the in-pixel CSA. 
This should be due to the reduction in the sense-node parasitic capacitance by making the area of the BNW smaller, 
and also to the increase in the closed-loop gain by reducing the feedback parasitic capacitance between the CSA and the BNW\cite{Miyoshi:2017fu,Miyoshi:2018cu,Takeda:2018xx}. 

\subsection{Pinned Depleted Diode structure}
In the single SOI structure, the charge generated in the interface region between the sensor and BOX layers is collected, 
which results in a significantly large dark current degrading the spectral performance. 
The device also suffers from the possibility of signal charge loss by the traps at the interface, 
which results in the degradation of the charge collection efficiency\cite{Matsumura:2015dq}. 
This situation is unchanged even with the Double-SOI structure. 

\begin{figure} [ht]
	\begin{center}
	\begin{tabular}{c} 
	\includegraphics[width=8cm]{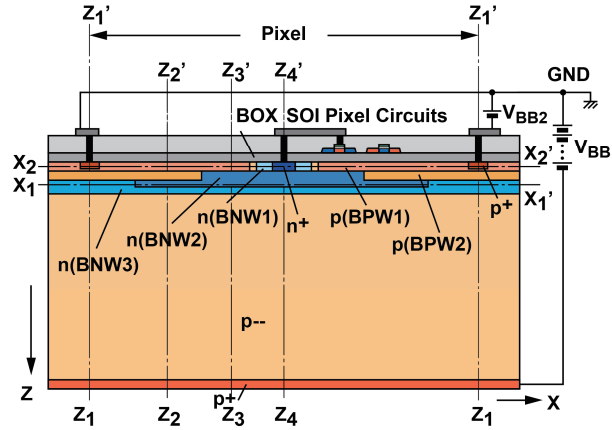}
	\end{tabular}
	\end{center}
	\caption[example] 
	{ \label{fig:2018Kamehama_Fig3} 
	Cross-sectional view of SOIPIX-PDD. The figure is adopted from Kamehama et~al. (2018)\cite{Kamehama:2017kp}.
	}
\end{figure}

In order to solve these problems, Kamehama et~al. (2018) recently developed a Pinned Depleted Diode (PDD) structure\cite{Kamehama:2017kp}. 
The PDD structure has a BPW region beneath the BOX layer, and a BNW region below that in the single SOI wafer as shown in Figure~\ref{fig:2018Kamehama_Fig3}. 
The signal charge generated by an X-ray collects through the stepped buried n-well (BNW1, BNW2 and BNW3) into the readout node (n$^+$) without touching the interface between the sensor (p--) and the BOX layers. 
Thus, the signal charge loss by the traps at the interface does not occur. 
The sufficiently highly doped buried p-well (BPW1) as a neutral region reduces the dark current generation from the interface.  
It works just like a pinned photodiode in CCD or CMOS image sensors. 
The buried p-well acts as a shielding layer between the sensor and circuit layers, and suppresses the interference between the two layers in the same way as the middle Si layer in the Double-SOI structure. 

\begin{figure} [ht]
	\begin{center}
	\begin{tabular}{c} 
	\includegraphics[width=7cm]{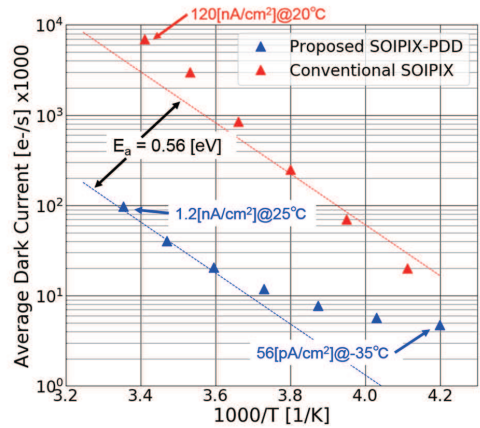}
	\end{tabular}
	\end{center}
	\caption[example] 
	{ \label{fig:2018Kamehama_Fig15} 
	Dark current per pixel as a function of working temperature in the single SOI device (Conventional SOIPIX) and 
	in the PDD device (Proposed SOIPIX-PDD). 
	The figure is adopted from Kamehama et~al. (2018)\cite{Kamehama:2017kp}. 
	}
\end{figure}

Figure~\ref{fig:2018Kamehama_Fig15} shows the dark current per pixel as a function of working temperature\cite{Kamehama:2017kp}. 
The dark current generation of the device with the PDD structure is two orders of magnitude lower than that of the device with the single SOI at $25{\rm ^{\circ}C}$. 
A dark current value of $10\ {\rm e\ msec^{-1}\ pixel^{-1}}$ can be achieved by cooling the device to below $-10{\rm ^{\circ}C}$. 
We note that the temperature dependency tends to saturate at a temperature lower than $-5{\rm ^{\circ}C}$. 
Kamehama et~al. (2018) suggest trap-assisted band-to-band tunneling as the possible reason for the saturation\cite{Kamehama:2017kp}. 

%
%

\begin{figure} [ht]
	\begin{center}
	\begin{tabular}{c} 
	\includegraphics[width=14cm]{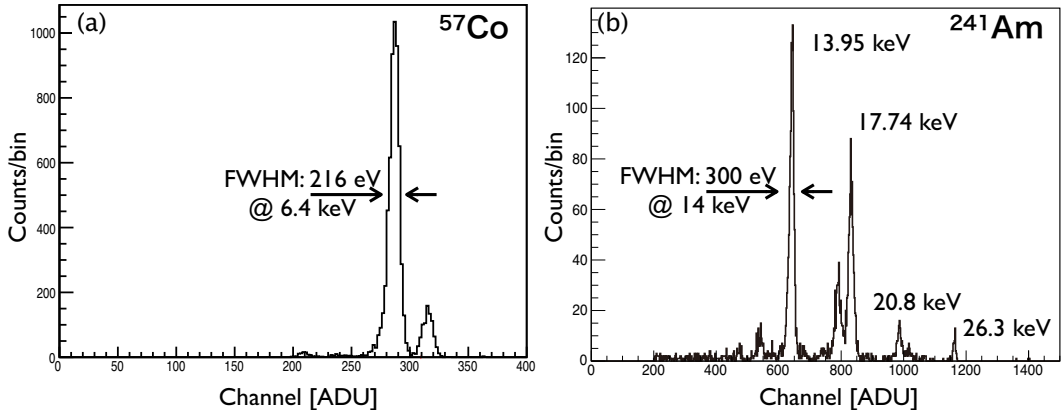}
	\end{tabular}
	\end{center}
	\caption[example] 
	{ \label{fig:XRPIX6E-PFZ-FI-200um_EventDriven_Spec} 
	X-ray spectra of Co-57 (left) and Am-241 (right) obtained in the Event-Driven readout mode with XRPIX6E\cite{Harada:2018xx}. }
\end{figure}

The device with the PDD structure exhibits a better spectral performance than that with the single SOI structure as well as that with the Double-SOI structure. 
Figure~\ref{fig:HistSpectralPerformance} (f) shows the X-ray spectrum of Fe-55 obtained with single pixel in the device (SOIPIX-PDD) having the PDD structure developed by Kamehama et~al. (2018)\cite{Kamehama:2017kp}. 
The energy resolution of $200~{\rm eV}$ in FWHM at 5.9~keV and the ENC of $11~{\rm e}$ in rms are the best values ever obtained with the X-ray SOI pixel sensors, although this device does not support the Event-Driven readout mode. 
Figure~\ref{fig:XRPIX6E-PFZ-FI-200um_EventDriven_Spec} shows the X-ray spectra of Co-57 and Am-241 obtained in the Event-Driven readout mode with a new device, XRPIX6E, which is the first PDD device supporting the Event-Driven readout mode\cite{Harada:2018xx}. 
While the spectra were collected from single arbitrarily chosen pixel, 
it was already confirmed that the device exhibits the same spectral performance when using other single pixel. 
The energy resolution is $216~{\rm eV}$ at $6.4~{\rm keV}$, which is close to that obtained in the Frame readout mode. 



\section{Summary and Future Prospects}
After about 10 years of development, we have successfully realized a device with a sensor size of $21.9~{\rm mm} \times 13.8~{\rm mm}$ and a format of $608\times 384$, and achieved a time resolution higher than $10~{\rm \mu sec}$ and a throughput higher than $570~{\rm Hz}$ in the Event-Driven readout mode. 
We also successfully reduced the interference between the sensor and circuit layers by improving the device structure. 
The latest device shows an energy resolution of $216~{\rm eV}$ in FWHM at 6.4~keV in the Event-Driven readout mode. 

In order to realize a high spectral performance and a large format at the same time, we are now developing a new device with an imaging area size of $21.9~{\rm mm} \times 13.8~{\rm mm}$ having the Double-SOI or PDD structure.  
While the energy resolution at or above 6~keV was significantly improved, 
the X-ray performance at a low X-ray energy band below 2~keV is still insufficient for the FORCE mission\cite{Itou:2016io}. 
Therefore, we will study the backside process using the latest devices, XRPIX6D and XRPIX6E, and make necessary improvements. 

\acknowledgments 
We acknowledge the valuable advice and great work by the personnel of LAPIS Semiconductor Co., Ltd. 
This study was supported by the Japan Society for the Promotion of Science (JSPS) KAKENHI Grant-in-Aid for Scientific Research on Innovative Areas 25109002 (Y.A.), 25109003 (S.K.), 25109004 (T.G.T., T.T., K.M. and A.T.), 20365505 (T.K), 23740199 (T.K), 18740110 (T.K.), Grant-in-Aid for Young Scientists (B) 15K17648 (A.T.), Grant-in-Aid for Challenging Exploratory Research 26610047 (T.G.T.) and Grant-in-Aid for JSPS Fellows 15J01842 (H.M.). 
This study was also supported by the VLSI Design and Education Center (VDEC), the University of Tokyo in collaboration with Cadence Design Systems, Inc., and Mentor Graphics, Inc.

\bibliography{RefPublished_v1,RefSubPrep} 
\bibliographystyle{spiebib} 

\end{document}